\documentclass[12pt]{iopart}
\usepackage{iopams}
\usepackage{graphicx,cite,epsf}
\usepackage{psfrag}
\usepackage{ulem}
\RequirePackage{color}
\definecolor{MyDarkGreen}{rgb}{0.02,0.60,0.06}

\begin{document}

\title[On the shape of Gaussian scale-free polymer networks]{On the shape of Gaussian scale-free polymer networks}

\author{V. Blavatska$^{1,2}$\footnote{Author to whom any correspondence should be addressed} and Yu. Holovatch$^{1,2,3,4}$}
\address{$^1$ Institute for Condensed Matter Physics of the National Academy of Sciences of Ukraine,
79011 Lviv, Ukraine}
\address{$^2$ $\mathbb{L}^4$ Collaboration $\&$ Doctoral College for the Statistical Physics of Complex Systems, Leipzig-Lorraine-Lviv-Coventry, Europe}
\address{$^3$ Institute of Mathematics, Maria Curie-Sk\l odowska University, 20-031 Lublin, Poland}
\address{$^4$ Complexity Science Hub Vienna, 1080 Vienna, Austria }
\ead{viktoria@icmp.lviv.ua}
\begin{abstract}

We consider the model of  complex hyperbranched polymer structures formed on the basis of scale-free graphs, where functionalities (degrees) $k$ of nodes obey a power-law decaying probability $p(k)\sim{k^{-\alpha}}$.
Such polymer topologies can be considered as generalization of regular hierarchical dendrimer structures with fixed functionalities.  
 The conformational size and shape characteristics, such as averaged asphericity
$\langle A_3 \rangle$ and size ratio $g$ of such polymer networks are obtained numerically by application of Wei's method, which defines the configurations of any complex Gaussian network in terms of eigenvalue spectra of the corresponding Kirchhoff matrix. Our quantitative results indicate, in particular,  an increase of compactness and symmetry of network structures with the decrease of parameter $\alpha$.  

\end{abstract}
\date{\today}
\submitto{Journal of Physics: Condensed Matter}
\maketitle

\section{Introduction}

\begin{figure}[h!]
	\begin{center}
		\includegraphics[width=70mm]{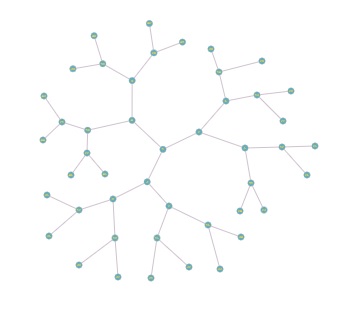}
\includegraphics[width=70mm]{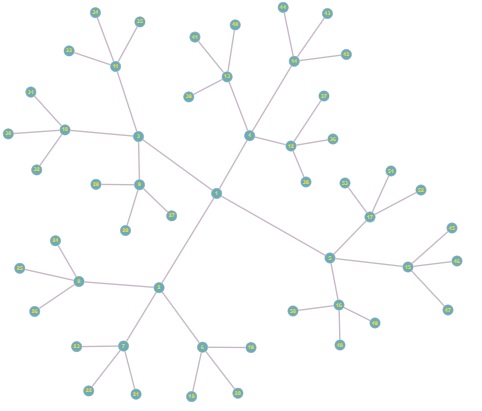}	
 \end{center}
\caption{Snapshots of dendrimers D(3,3) and D(4,2) constructed on a basis of Cayley tree algorithms.}
	\label{snapdend}
\end{figure}

The impact of topology of individual polymer macromolecules on their static and dynamic properties in solvents is a question of great importance in modern polymer physics. Starting with the seminal works of Zimm and Stockmayer \cite{Zimm1949}, a theoretical analysis of the simplest representative of a non-trivial macromolecule topology, so-called star polymer  (see e.g \cite{CM})
with a single branching point has been performed.
With the continuous progress in polymer synthesis and analysis, the  macromolecules  with very complex multiply branched architectures and tunable properties have been synthesized. 
In particular, the topology of many synthetic highly branched molecules can
 be successfully represented by an idealized model of finite Cayley tree structure,
 governed by iterative
branching process. Such structures, known as dendrimers \cite{Tomalia1985} or cascade macromolecules \cite{Buhleier1978} are composed of a central core, branched units called ”generations,” and terminal
functional groups. An initiator core, considered as the first generation $n=0$, is
presented as a node with $k$ outgoing links with  ``periphery'' nodes attached, each of them  thus serving as an initiator for  successive $n>0$th generation (Fig. \ref{snapdend}). Note, that $0$th generation of a dendrimer structure thus recovers the star-like topology. A comprehensive and recent review discussing the properties and
potential applications of dendrimers across different fields of research,
technology and treatment can be found in Ref. \cite{Mathur10}.  Many theoretical studies are using an idealized Gaussian approximation  to analyze the properties of complex polymers such as 
dendrimers \cite{Cai1997,Chen1999,Ferla1997,Gurtovenko2003,Galiceanu2007,Galiceanu2010,deRegt2019,Jura2021},  star-based
topologies \cite{Biswas2000,Satmarel2005,Poliska2014,Ganazzoli2002}, comb polymers \cite{Ferber2015,deRegt2016,Bishop2017}, polymer brushes \cite{Haydukivska2022}, tree-like structures \cite{Dolgushev2017}, snowflake polymers \cite{Haydukivska2023}, Erd\H{o}s-R{\'e}nyi  polymer networks \cite{Blavatska2020},  multihierarchichal  structures constructed on the basis of Sierpinski gasket \cite{Jurjiu2017}, 
Vicsek fractal \cite{Jurjiu2017a} 
 T-fractal \cite{Mielke2016} etc.

The polymer networks with very high (classically referred to as “infinite”) molecular weights are often encountered in materials like  synthetic and biological gels, thermoplastics, elastomers  \cite{Clark87,Burchard1990,Gu2020}. Polymer networks are the interesting objects to study  both from
  academic \cite{Duplantier1989,Safer1992} and commercial \cite{Gao2004,Jeon2018,Gu2020} perspective.
  In the mathematical language of  graphs, the branching points in networks are considered as
vertices (nodes), and their functionalities thus define the number of outgoing links (degrees) of these nodes.  In particular, in Ref. \cite{Blavatska2020}  the model of a random polymer network, formed on the base on  Erd\H{o}s-R{\'e}nyi random graph was proposed.
  In Refs. \cite{Galiceanu2014,Galiceanu2012}, the scale-free polymer networks have been constructed and their dynamical properties analyzed on the basis of the Rouse model.  
In this case, the functionalities $k$ (the degrees) of nodes obey a power-law decaying  distribution \cite{Barabasi1999}
\begin{equation}
p(k)\sim{k^{-\alpha}},\,\,\,\,k\gg 1.  \label{dis}
\end{equation}
The dynamics of such polymer networks demonstrates a transition from predominant branched-like network behaviour (low values of 
$\alpha$) to linear-like networks
(high $\alpha$) \cite{Galiceanu2014}. 
{   The polyfunctional polymer network obeying the distribution (\ref{dis}) for functionalities of junction points has been synthesized recently in \cite{Pleshakov,Pleshakov2005} based on transformer oil plasticized butyl
rubber. To obtain the polyfunctional network, part of the
transformer oil was replaced by a blend of butadiene
oligomers  which functioned as polyfunctional cross-links.
 The mechanical properties of such networks have been analyzed and the high reliability and
stability of polyfunctional networks towards the external stress or damage was confirmed, thus 
 justifying the expedience of such structures in the manufacture of polymer materials.
}

\begin{figure}[h!]
	\begin{center}
		\includegraphics[width=70mm]{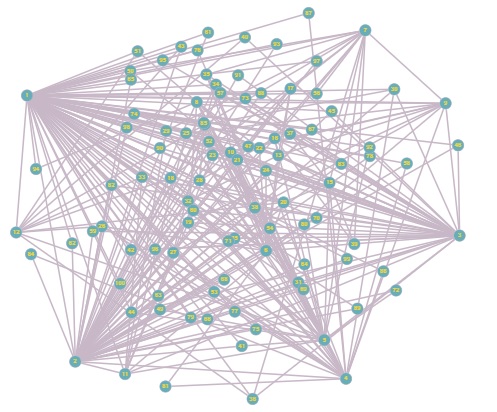}
\includegraphics[width=70mm]{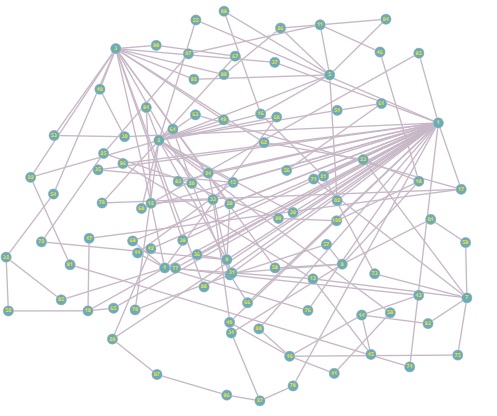}	
 \end{center}
\caption{Snapshots 
of scale-free networks with $N=100$ monomers with $\alpha=2.1$ (left) and $\alpha=3.5$ (right), constructed on a basis of the scale-free configuration model (SFC) algorithm with $k_{{\rm min}}=2$.}
	\label{snapnetfull}
\end{figure}

\begin{figure}[h!]
	\begin{center}
		\includegraphics[width=70mm]{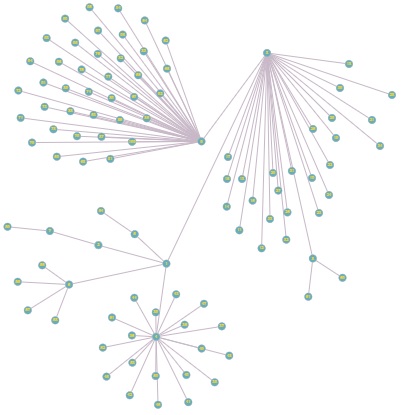}
\includegraphics[width=70mm]{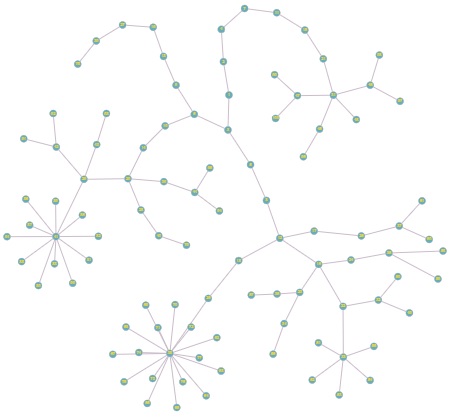}	
 \end{center}
\caption{Snapshots of scale-free networks with $N=100$ monomers with $\alpha=2.1$ (left) and $\alpha=3.5$ (right) constructed on a basis of the scale-free dendrimer (SFD) algorithm with $k_{{\rm min}}=1$.}
	\label{snapnet}
\end{figure}

The question of great interest is the effective size and shape characteristics of typical conformations, formed dy individual polymer macromolecules in solvents. Indeed, the hydrodynamics of polymer solutions
is affected by the size and shape of individual macromolecules \cite{Torre2001}, the shape of proteins influences their dynamics and folding   \cite{Plaxco1998,Ouyang2008} etc. 
In the pioneering work of Kuhn \cite{Kuhn1934} it was analytically predicted, based on combinatorial analysis considerations, that the Gaussian polymer chains
in solvents form the crumpled anisotropic coils, which can be rather approximated by prolate ellipsoids than spheres. The experimental studies of the viscous properties of polymer solutions confirmed that prediction of Kuhn \cite{Herzog1933,Perrin1936}.

{    Solc and Stockmayer \cite{Solc1} proposed to consider the set of normalized average eigenvalues $\lambda_i$ of the gyration tensor  as a measure of shape of polymer structures (measure of dispersion of monomers as related to the center of mass of macromolecule). In particular,  their  numerical simulations of the linear polymer chain in $d=3$ gave $\lambda_1\gg \lambda_2 \approx \lambda_3$\cite{Solc1}, indicating a high anisotropy of typical polymer configurations compared with the symmetric  case $\lambda_1=\lambda_2=\lambda_3$ (when monomers are uniformly distributed around the center of mass). 
In Ref. \cite{Aronovitz86,Rudnick86} it was proposed  to describe and classify the shape properties of polymer structures of any topology in $d$-dimensional space in terms of  averaged asphericity $A_d$. This quantity is defined as the variance of eigenvalues $\lambda_i$ divided by their squared mean, and is thus related to the index of dispersion. 
 This quantity is bounded by $ A_d=0$ for completely spherical configurations (uniform distribution around center of mass) and attains the maximal value $ A_d=1$ for completely anisotropic rod-like structure (maximally non-homogeneous case).}  
 Otherwise, its values $ 0<A <1 $  correspond to configurations with ellipsoid-like shape, and the larger is the value,
 the more extended and asymmetric  is the ellipsoid.
An effective size of any polymer structure  can be characterized by the mean-square
radius of gyration $\langle R_g^2 \rangle$.  To compare the size measures of any complex polymer topology (network) $\langle R_g^2 \rangle_{\rm network}$
and that of the linear polymer chain $\langle R_g^2 \rangle_{\rm chain}$ of the same total molecular weight (same number of monomers), their ratio 
{   (the so-called $g$-ratio) was introduced \cite{Zimm1949}:
\begin{equation}
g=\frac{\langle R_g^2 \rangle_{\rm network}}{\langle R_g^2 \rangle_{\rm chain}}. \label{gratio}
\end{equation}}
Both $A_d$ and $g$ values are universal, i.e. they are impacted only by topology  of a complex polymer and space dimension,
but not by  details of its chemical structure.
A considerable attention has been paid
to analytical and numerical analysis of $A_d$ and $g$ of the  macromolecules of linear \cite{Bishop1988,Diehl1989,Gaspari1987,Cannon1991,Sciutto1994,Haber2000},
circular \cite{Bishop1988,Diehl1989,Gaspari1987,Cannon1991,Jagodzinski1992}, and branched \cite{Bishop1993,Wei1997,Casassa1966,Ferber2013,Ferber2015,Blavatska2015,Haydukivska2020, Haydukivska2022,Haydukivska2023}  architectures. In  Ref. \cite{Blavatska2020} the size and shape characteristics of the set of Erd\H{o}s-R{\'e}nyi polymer networks have been evaluated.

In the present study, we suggest a powerful way to  analyze the size and shape characteristics of scale-free polymer networks numerically. To this end, we will apply the Wei's algorithm, which defines the network shape in terms of its Kirchhoff matrix. Being interested in impact of nodes of  high functionality, we will provide algorithms to construct the scale-free Gaussian networks (as shown in Figs. \ref{snapnetfull}, \ref{snapnet}) and apply numerical analysis to evaluate the eigenvalue spectra of their Kirchhof matrices.     The layout of the rest of the paper is as follows. In the next 
section, we give the definitions of the observable characteristics and provide the details of Wei's method, developed to evaluate them. In Section \ref{III}, we describe the algorithms applied to construct the scale-free polymer networks and discuss the results obtained. We end up with giving conclusions in Section \ref{IV}.  

 \section{Definitions and Wei's method}{\label{II}}
\subsection{Polymer size and shape characteristics}\label{IIa}
The size and shape characteristics of a polymer of any topology containing $N$ monomers (nodes or vertices in the language of graph theory) and embedded in $d$-dimensional space
can be characterized
  in terms of the gyration tensor $\bf{Q}$
with components:
\begin{equation}
Q_{ij}=\frac{1}{N}\sum_{m=1}^N(x_m^i-{x^i_{CM}})(x_m^j-{x^j_{CM}}),\,\,\,\,\,\,i,j=1,\ldots,d,
\label{mom}
\end{equation}
with $x_m^i$ being the $i$th coordinate of the position vector $\vec{R}_m$ of the $m$th monomer and   ${x^i_{CM}}=\sum_{m=1}^N x_n^i/N$ the $i$th 
coordinate of the center-of-mass position vector ${\vec{R}_{CM}}$.

 The squared radius of gyration $R_g^2$ is thus
   defined as a trace of the gyration tensor $\bf{Q}$:
\begin{equation}
R_g^2 ={\rm Tr}\, {\bf{Q}} =  \sum_{i=1}^d Q_{ii} =\frac{1}{N}\sum_{m=1}^N (\vec{R}_m-{\vec{R}_{CM}})^2. \label{rg1}	
\end{equation}

The set of eigenvalues $\lambda_i$ of the gyration tensor (\ref{mom}) directly characterizes the  degree  of asymmetry of  an instantaneous polymer configuration. For completely symmetric (spherical)
configuration, corresponding to uniform distribution of monomers around the center of mass, all the eigenvalues $\lambda_{i}$ are equal, whereas for the maximally anisotropic case (the stretched rod-like structure) all $\lambda_{i}$ are zero except one.
 The asphericity $A_d$,  connected with the  index of dispersion of the eigenvalue set, is then defined by 
 \cite{Aronovitz86}:
 \begin{equation}
 A_d=\frac{1}{d(d-1)}\sum_{i=1}^d\frac{(\lambda_i-{{\lambda}})^2}{{{\lambda}^2}}, \label{asfer}
\end{equation}
 where ${{\lambda}}\equiv \sum_{i=1}^d \lambda_i/d$ is a mean arithmetic of $\lambda_i$.  

Note that the averaging of any observable $O$  of interest has to be performed over an ensemble of $M$ constructed configurations of a macromolecule according to:
\begin{equation}
\langle O \rangle =\frac{1}{M}\sum_{m=1}^MO_m,\label{aver}
\end{equation}
where $O_m$ is the value of the observable $O$ estimated for the $m$th configuration.

In graph theory settings, the linear polymer chain of $N$ monomers can be considered as a network, where functionalities of two side monomers (nodes) are equal to 1, whereas all the $N-2$ inner nodes are sequentially connected to two of their neighbours and thus have degree 2. The circular polymer chain with both end monomer bonded together, can be described as a network with degrees of all nodes equal to two. In the following, we will be interested in Gaussian networks, neglecting 
self- and mutual-avoiding interactions between  monomers.  
For the Gaussian polymer structure in a form of closed ring the exact value of the size ratio is evaluated \cite{Zimm1949}:
\begin{equation}
g_{\rm ring}=\frac{1}{2}. \label{gratioring}
\end{equation}

Note  that two essential possibilities arise in considering the branched polymer structures.
In the first case,  a considerable
number of monomers (nodes)  are linked in linear sequence between any two branching point, building up so-called polymer strands.
In such view, e.g. the star polymer of functionality $k$ is considered as a set of linear chains of length (number of nodes) $N$, connected to a central node with degree $k$.  
In the second case, the segments between any two branching points are treated just as a link between two nodes (a direct bond). In this view, e.g. for the Gaussian dendrimer structures with functionality $k$ and generation $n$ the size ratio was evaluated exactly \cite{Ferla1997}:
\begin{eqnarray}
&&g_{{\rm dendrimer}}(k,n)=6k\Big\{(k-1)^{2(n+1)}\Big[k(n+1)(k-1)-k(n+1)\nonumber\\
&&-2k+1\Big]+(k-1)^{n+1}\Big[2k+k(n+1)-k(n+1)(k-1)-n-1\nonumber\\
&&+(n+1)(k-1)^2\Big]-1\Big\}/\Big(k-2+k\Big[(k-1)^{n+1}-1\Big]\Big)^3. \label{gdend}
\end{eqnarray}
The same view was exploited when studying the scale-free polymer networks in Refs. \cite{Galiceanu2007,Galiceanu2014}, and we will concentrate on the same picture in what follows.

\subsection{Wei's method}\label{IIb}

To find the quantitative estimates for the observables (\ref{gratio}) and (\ref{asfer}) for polymer network of a given topology in Gaussian approximation we will apply the powerful  Wei's method \cite{Wei,Ferber2015}. This method defines the shape of a Gaussian network of $N$ vertices in terms of its $N\times N$ Kirchhoff matrix  $\bf{K}$. 
{    Note that   the Kirchhoff matrix is also known as the combinatorial Laplacian in graph theory. It is  defined as $\bf{K}=\bf{D}-\bf{A}$, with ${\bf D}$ being the degree matrix and ${\bf A}$  the adjacency matrix.}
 Thus, the diagonal elements $K_{ii}$ equal
the degree of vertex $i$, whereas the non-diagonal elements $K_{ij}$ equal  $-1$ when the vertices $i$ and $j$ are adjacent and $0$ otherwise. E.g., for the most trivial network corresponding to a linear chain of $N$ monomers as mentioned above, one has $K_{jj}=2$ for $j=2,\ldots,N-1$, $K_{11}=K_{NN}=1$, and $K_{j\,j+1}=K_{j\,j-1}=-1$, the rest of the matrix elements being zero.
Let $\sigma_2,\ldots,\sigma_N$ be $(N-1)$ non-zero eigenvalues of the $N\times N$
Kirchhoff matrix ($\lambda_1$ is always $0$). 

{    The matrix ${\bf L}(y)={\bf E}-y$ is introduced next, where  ${\bf E}$ is $(N-1)\times(N-1)$ diagonal matrix whose diagonal elements are the eigenvalues of ${\bf K}$ (excluding the zero eigenvalue), $y$ is an  auxiliary variable.  The
 functions needed for the calculation of the shape parameters are defined next:
\begin{eqnarray}
&&G(y)={\rm Det}({\bf L}^{-1}(0)){\rm Det}({\bf L}(y^2))=\prod_{j=1}^{N-1}\sigma_j^{-1}\prod_{j=1}^{N-1}(\sigma_j+y^2),\\
&& S_1(y)=\frac{1}{N^2}{\rm Tr} {\bf L}^{-1}(y^2)=\frac{1}{N^2}\prod_{j=1}^{N-1}(\sigma_j+y^2)^{-1},\\
&&S_2(y)=\frac{1}{N^2}{\rm Tr} {\bf L}^{-2}(y^2)=\frac{1}{N^2}\prod_{j=1}^{N-1}(\sigma_j+y^2)^{-2}.
\end{eqnarray}

{   Note that $S_1(0)$ restores the so-called Kirchhoff
index (a structure-descriptor), which is connected with 
the set of resistance distances $r_{ij}$ in a graph via 
\begin{equation}
S_1(0)=N\sum_{i<j}r_{ij}.
\end{equation}
The  resistance distance $r_{ij}$ between nodes $i$ and $j$ on a graph is a distance function on a graph, which characterizes the multiple-route distance diminishment, while considering any link between vertices as effective resistance  \cite{Klein1993,Chen2007}. 
}

The $g$-ratio can thus be defined by
\begin{equation}
g=\frac{S_1^{{\rm network}}(0)}{S_1^{{\rm chain}}(0)}=\frac{\sum_{j=2}^{N}1/\sigma_j^{{\rm network}}}{\sum_{j=2}^{N}1/\sigma_j^{{\rm chain}}}\, ,
\label{gwei}
\end{equation}
where $\sigma_j^{{\rm network}}$ and $\sigma_j^{{\rm chain}}$ are the Kirchhoff matrix eigenvalues for the polymer 
network and a linear chain of the same number of monomers, correspondingly.
In turn, the asphericity of the corresponding Gaussian polymer structure in $d$ dimensions can be obtained by \cite{Wei,Ferber2015}:
\begin{eqnarray}
&&\langle A_d \rangle =\frac{d(d+2)}{2}\int_0^{\infty}x^3G^{-d/2}(x)S_2(x){\rm d}x=\nonumber\\
&&=\frac{d(d+2)}{2}\int_0^{\infty} {\rm d} y \sum_{j=2}^{N}\frac{y^3}{(\sigma_j+y^2)^2}\left[ \prod_{k=2}^{N} 
\frac{\sigma_k}{\sigma_k+y^2}\right ]^{d/2},
\label{awei}\end{eqnarray}

}
 \begin{figure}[b!]
	\begin{center}
		\includegraphics[width=100mm]{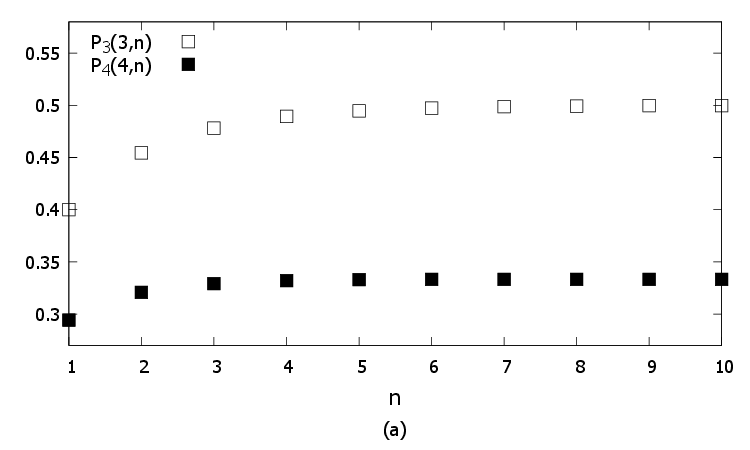}
		\includegraphics[width=100mm]{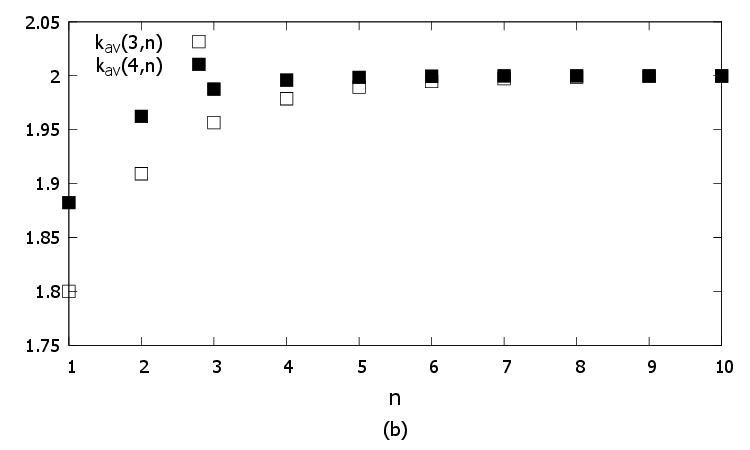}
	\end{center}
	\caption{ Probabilities to obtain a node of degree $k$ (a) and the averaged node degree $k_{av}(k,n)$(b) as functions of $n$ for two dendrimer structures:  D($k=3$,n)  (open symbols) and D($k=4$,n)  (filled symbols).}
	\label{probdend}
\end{figure}

The Wei's method has been successfully applied to evaluate the shape values for the set of complex polymer structures such as comb polymers \cite{Ferber2015,deRegt2016,Bishop2017}, dendrimers \cite{deRegt2019,Jura2021}, polymer brushes \cite{Haydukivska2022}, tree-like structures \cite{Dolgushev2017} snowflake polymers \cite{Haydukivska2023}, Erd\H{o}s-R{\'e}nyi polymer networks \cite{Blavatska2020}. Below, we will apply it in the scale-free network analysis.

\section{Network construction algorithms and Results}\label{III}

\subsection{Regular dendrimer structures}\label{IIIa}
Let us start with 
reconsidering the regular Gaussian dendrimer structures, which are uniquely determined by the set of two variables  $k$ (functionality) and $n$ (the generation) and will be defined as D(k,n) in what follows. Two examples of such structures, namely D(3,3) and D(4,2) are shown in Fig. \ref{snapdend}.

 The total number of nodes $N(k,n)$ in such structures  and the number of ``periphery'' nodes in the outer $n$th generation shell $N_{out}(k,n)$ can be easily calculated as: 
\begin{eqnarray}
 &&  N(k,n)=1+\frac{k((k-1)^{n+1}-1)}{k-2}\, , \\
 && N_{out}(k,n)=k(k-1)^n=\frac{N(k-2)+2}{k-1} \, .
\end{eqnarray}
Note that all nodes in the outer shell have the degree 1, whereas the inner $N(k,n)-N_{out}(k,n)$ nodes are branching points of functionality $k$. We can thus introduce the probability $P_k(k,n)=(N(k,n)-N_{out}(k,n))/N(k,n)$ for $k$-branching node to occur in $n$th generation, which is depicted in Fig. \ref{probdend}a for two denrimers D(3,n) and D(4,n) with $k=3$ and $k=4$, correspondingly. This probability reaches a constant value with increasing $n$. 
The average node degree in dendrimer structure can thus be estimated as:
{   
\begin{eqnarray}
  &&  k_{av}=\frac{N_{out}(k,n)+k(N(k,n)-N_{out}(k,n))}{N(k,n)}=\nonumber\\
  && = (1-k)\frac{N_{out}}{N}+k=2-\frac{2}{N}. \label{kavdend}
\end{eqnarray}
which tends to the value $k_{av}=2$ in the limit $N\to\infty$ (see Fig. \ref{probdend}b).
 This estimate for the average node degree of tree-like structures is independent of their 
functionalities. It has been found previously in Ref. \cite{Katzav18} based on 
different considerations. 
}

Before going any further, let us note that within the framework of Wei's method, the peculiarities of the Kirchhoff matrix eigenvalue spectrum $\{\sigma\}$   are determinative in evaluating the size and shape conformation characteristics. The eigenvalue spectra of denrimers D(3,$n$) up to $n=6$ were evaluated exactly in \cite{Cai1997}, and for general tree-like structures in \cite{Dolgushev2017}.

\begin{figure}[h!]
	\begin{center}
		\includegraphics[width=100mm]{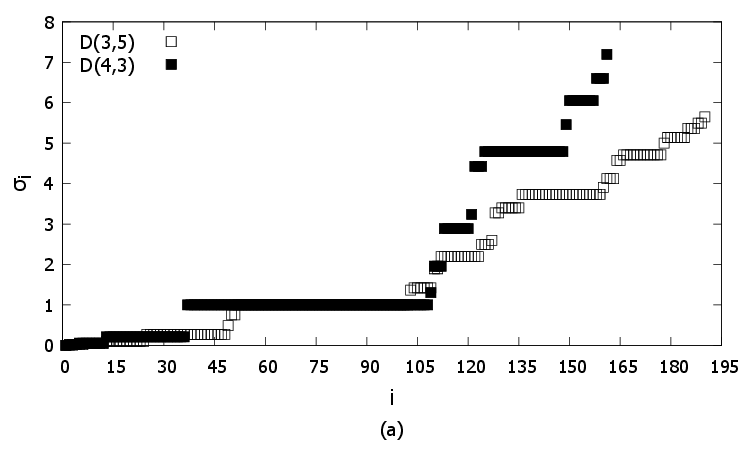}
		\includegraphics[width=100mm]{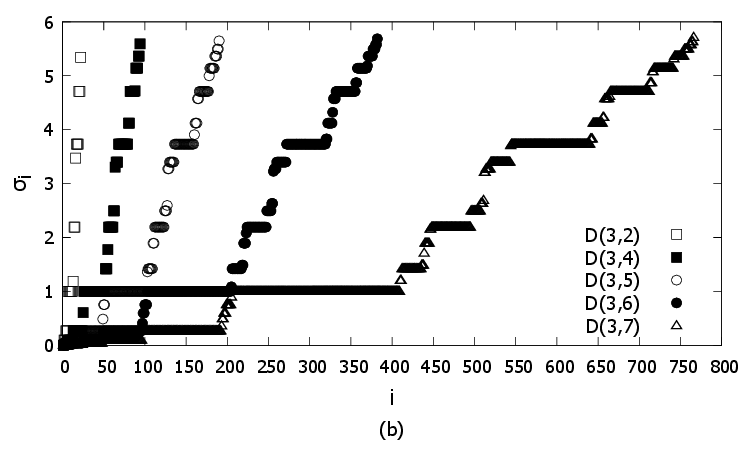}
	\end{center}
	\caption{The eigenvalue spectra of Kirchhoff matrix of dendrimer structures D(3,5), D(4,3) (a) and D(3,n) at  $n=2,4,5,6,7$ (b).  }
	\label{EigenBethe}
\end{figure}

  An estimate for an upper bound $\sigma_{{\rm sup}}$  for the eigenvalue spectrum of Kirchhoff matrix of a graph containing $i=1,\ldots,N$ nodes of degrees $k_1,\ldots,k_N$ correspondingly has been found in  \cite{Guo2005}:
\begin{equation}
\sigma_{{\rm sup}}\leq {\rm max}\left\{ \frac{k_i+\sqrt{k_i^2+8k_im_i}}{2}\right\},\,\,m_i=\sum_{j}(k_i-|N_i \cup N_j|)/k_j, \label{maxeig}
\end{equation}
with $|N_i \cup N_j|$ being the number of common neighbors of nodes $i$ and $j$. On the other hand, according to  the definition (\ref{gwei}) for the size ratio, the complex polymer structure is expected to be more compact ($g$ is smaller), the larger is the maximal eigenvalue. Therefore, an estimate from above of the spectra of Kirchhoff matrix eigenvalues allows one to evaluate the polymer network compactness.

For the case of a graph in a form of a linear chain, the upper limit  $\sigma_{{\rm sup}}^{{\rm chain}}=4$ can thus be obtained (with $k_i=2$, $k_j=1$), whereas for D(3,n) and D(4,n) dendrimers one has correspondingly $\sigma_{{\rm sup}}^{{\rm D(3,n)}}=6$ and $\sigma_{{\rm sup}}^{{\rm D(4,n)}}=8$. In Fig. \ref{EigenBethe}a we present the set of eigenvalues for the Kirchhoff matrix calculated for D(3,5) and D(4,3) dendrimers (we chose these two examples, since the number of nodes in resulting structures and, correspondingly, the number of eigenvalues are comparable, 190 and 161, correspondingly). Since the eigenvalues of  the Kirchhoff matrix of D(4,3) take on  larger values comparing to D(3,5), the former structure is expected to be more compact and symmetric comparing with the latter. The dependence of eigenvalue spectrum of $D(3,n)$ on the number of generations is depicted in Fig. \ref{EigenBethe}b.

\begin{figure}[h!]
	\begin{center}
		\includegraphics[width=100mm]{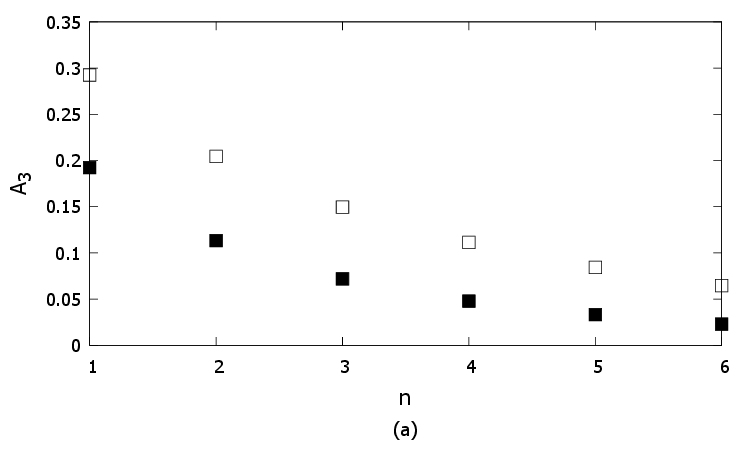}
		\includegraphics[width=100mm]{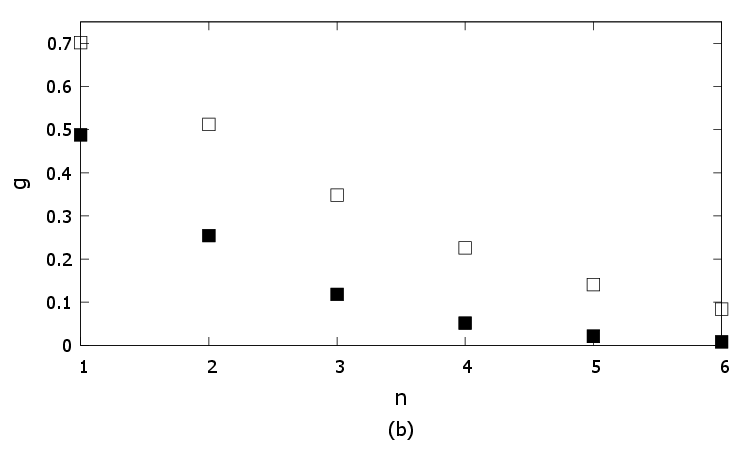}
	\end{center}
	\caption{Asphericities $ A_3$ (a) and size ratios $g$ (b) of dendrimer structures, obtained in numerical simulations with application of Wei formulas (\ref{awei}) and (\ref{gwei}). Open symbols: D(3,n),  filled symbols D(4,n). Note that the results for $g$ obtained via Wei's method perfectly coincide with those given by Eq. (\ref{gdend}). }
	\label{ABethe}
\end{figure}

We applied the Wei's method next to evaluate the asphericity $A_3$ and size ratio $g$ of  
two structures $D(3,n)$ and $D(4,n)$. Results are presented in Fig. \ref{ABethe} as functions of generation $n$. For the size ratios $g$, our numerical data perfectly coincide with analytical result as given by Eq. (\ref{gdend}).   Both $A_3$ and $g$ values tend to zero with increasing $n$, and thus with increasing the total size $N(k,n)$ of the dendrimer structure. {    This tendency towards the uniform spherical distribution of nodes with increasing the size of trees has been observed also in Ref. \cite{Budnick23}.} Though the averaged node degree $k_{av}(k,n)$ tends to the value of two like for a regular chain topology, the constant high probability to obtain the branching node (cf. Fig. \ref{probdend}a) leads to increasing the number of such nodes with growing of $N(k,n)$, resulting in compactification of such structure as compared with linear chain of the same total size.

\begin{figure}[b!]
	\begin{center}
	\includegraphics[width=100mm]{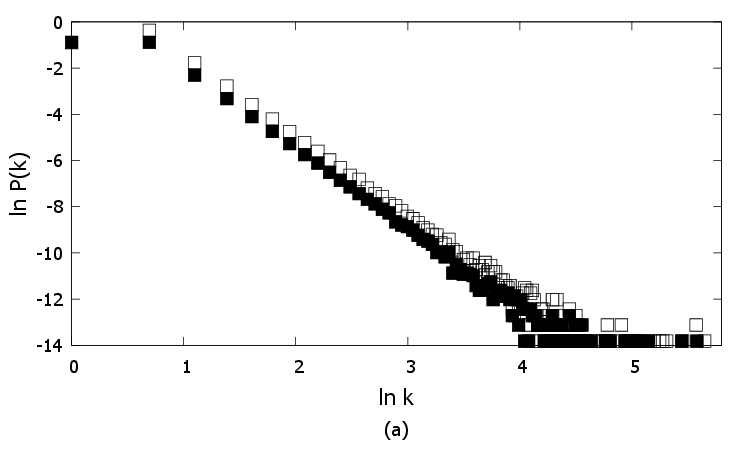}	
  \includegraphics[width=100mm]{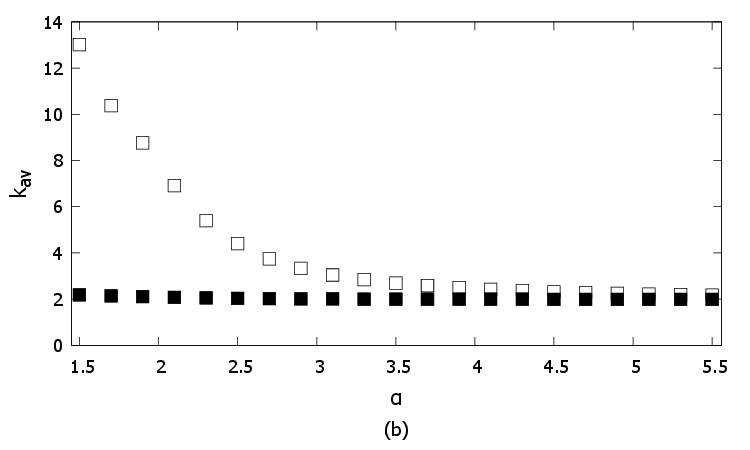}
 \end{center}
\caption{The node-degree distributions at $\alpha=3.5$ (a) and the averaged node degree $k_{av}$ as function of parameter $\alpha$ (b) of $N=100$ networks. Open symbols: SFC algorithm; filled symbols: SFD algorithm. }
	\label{kser}
\end{figure}

Next, let us construct the more complex polymer structures with node degrees distributed according to (\ref{dis}). Doing so, we will end up with two different structure topologies, although being 
governed by a similar node-degree distribution (\ref{dis}), they are characterized by different shape properties as will be shown below. 

\begin{figure}[t!]
	\begin{center}
		\includegraphics[width=100mm]{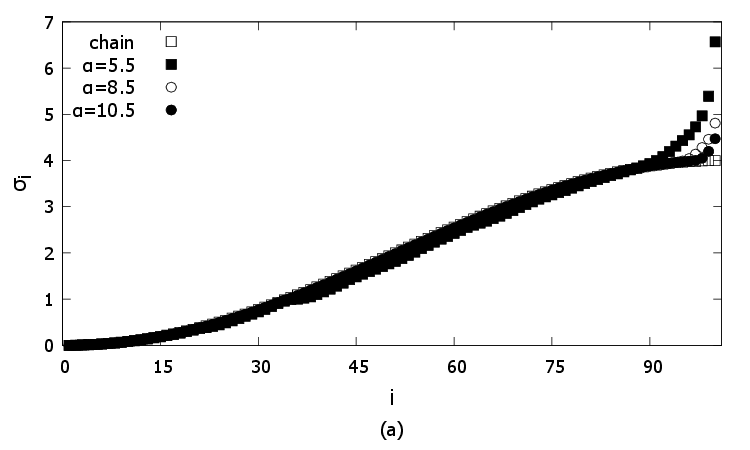}
		\includegraphics[width=100mm]{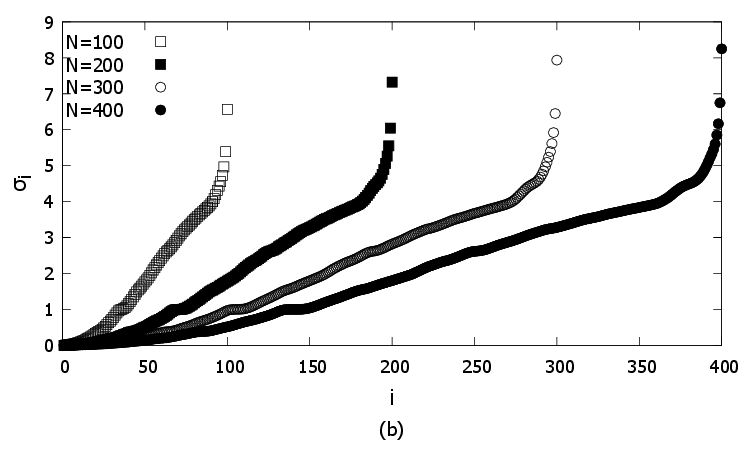}
	\end{center}
	\caption{The eigenvalues spectra of Kirchhoff matrix of scale-free network at several values of $\alpha$ (a) and at fixed $\alpha=5.5$ and several values of $N$ (b)  }
	\label{EigenNet}
\end{figure}

\begin{figure}[t!]
	\begin{center}
		\includegraphics[width=100mm]{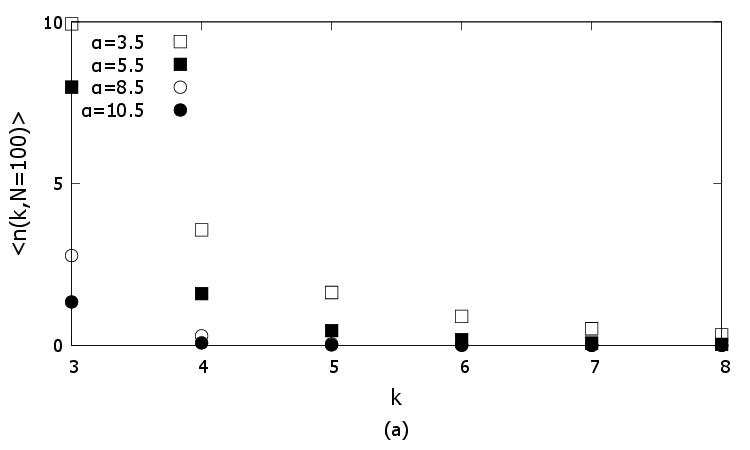}
		\includegraphics[width=100mm]{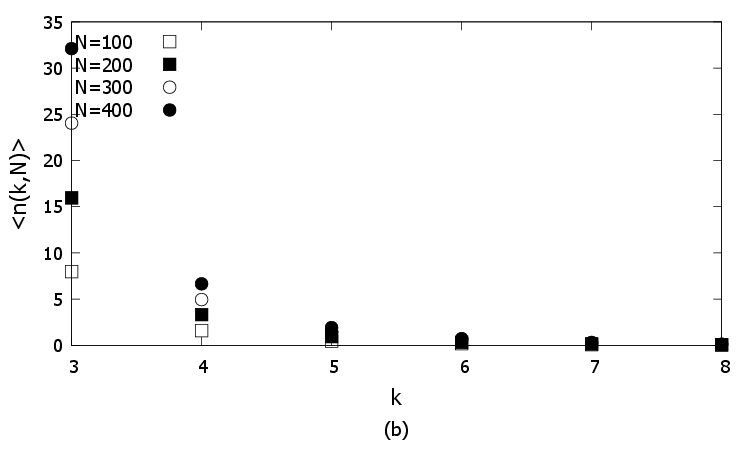}
	\end{center}
	\caption{The averaged numbers $\langle n(k,N) \rangle$ of nodes with degree $k$  at $N=100$ and several values of $\alpha$ (a) and at fixed $\alpha=5.5$ and several values of $N$ (b). }
	\label{nnet}
\end{figure}

\subsection{Scale-free configuration model (SFC) algorithm}\label{IIIb}

Following the scheme of configuration
model  algorithm \cite{Bender1978,Molloy1995}, we start with a set of $N$ disconnected nodes. The degrees $k_i$ ($i=1,\ldots,N$) are assigned to each of the nodes, as the random numbers selected from the probability distribution (\ref{dis}) with a condition $k_{{\rm min}}\leq k_i \leq k_{{\rm max}}$. Here,  $k_{{\rm min}} $ and $k_{{\rm max}}$ are desired minimal and maximal degrees. We take $k_{{\rm min}}=2$  to provide connectivity of network (the condition of existence of  giant connected component)  \cite{Molloy1998,Cohen2000}. 
We introduce the maximal degree cutoff $k_{{\rm max}}\sim N ^{1/2}$ according to   \cite{Catanzaro2005} in order to decrease the degree correlations in the configuration model.
Finally, the network is constructed by randomly connecting the nodes, taking into {account} the prescribed numbers of their outgoing links $k_i$ and controlling the avoidance of multiple connections and self-connections.  {   Such algorithm was applied for construction of scale-free networks in our previous paper \cite{Blavatska2022epid}}. Typical networks constructed on the basis of this algorithm are presented in Fig. \ref{snapnetfull}. 
Note that the networks created on the basis of this algorithm contain multiple loops, while as a model for 
hyperbranched polymers it is more useful to consider tree-like (dendrimer-like) structures. For this reason, a modified version of the standard scale-free network {generation} algorithm has been developed in  Refs. \cite{Galiceanu2007,Galiceanu2012,Galiceanu2014}, which is described next.

\begin{figure}[t!]
	\begin{center}
		\includegraphics[width=100mm]{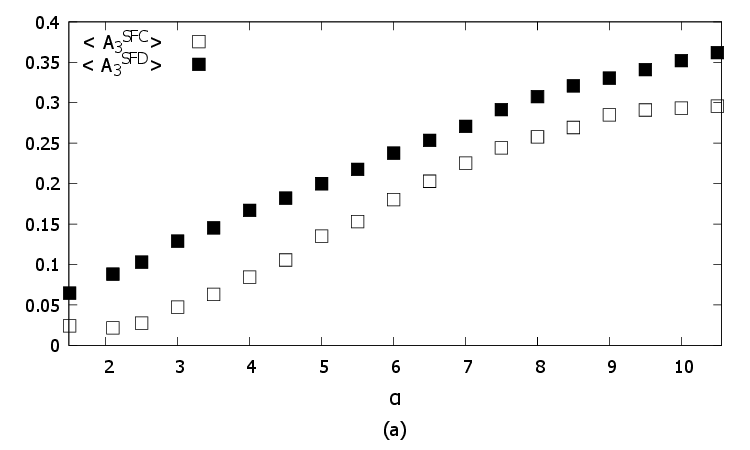}
		\includegraphics[width=100mm]{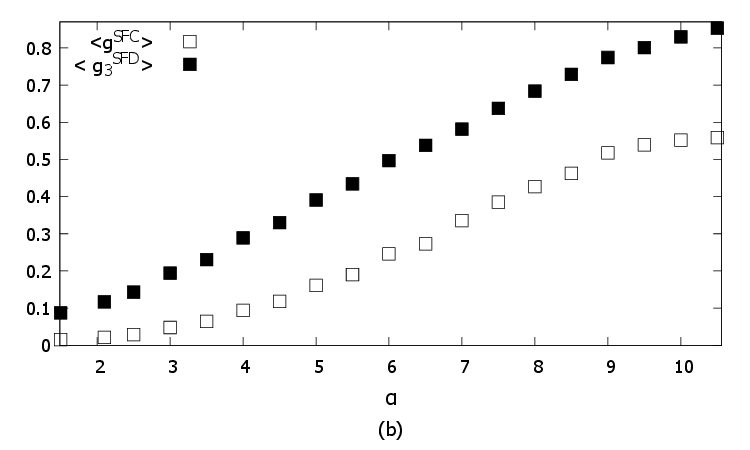}
	\end{center}
	\caption{Asphericities $\langle A_3\rangle$ (a) and size ratios $\langle g \rangle$ (b) of polymer scale-free networks with fixed $N=100$ and various $\alpha$ obtained in numerical simulations with application of Wei formula (\ref{gwei}). Open symbols: SFC algorithm; filled symbols: SFD algorithm. }
	\label{A100}
\end{figure}

\subsection{Scale-free dendrimer (SFD) algorithm}\label{IIIc}

Within this algorithm, we start with a single node and pick its degree $k_1$
randomly according to the distribution (\ref{dis}) with condition $k_{min}=2$.
The $k_1$ nodes are thus connected to each of its outgoing links.
Then, for each of these open nodes we again chose their degrees according to (\ref{dis}) and connect new open nodes to their outgoing links. The procedure is iterated by picking open
vertices and prescribing degrees to them, until the total desired number of nodes $N$ is reached. At this stage, the network  growth procedure is stopped and the degree $k=1$ is assigned to  all the remaining open
vertices.  In this method, the probability
to create a node with degree one is zero during the network growth process, but when the process stops the
networks will contain the
peripheral nodes with degree one.

The examples of networks obtained on the basis of this algorithm are presented in Fig. \ref{snapnet}, they have a typical topology of tree-like hierarchical structures and resemble the regular dendrimers, cf. Fig. \ref{snapdend}.

\begin{figure}[t!]
	\begin{center}
		\includegraphics[width=100mm]{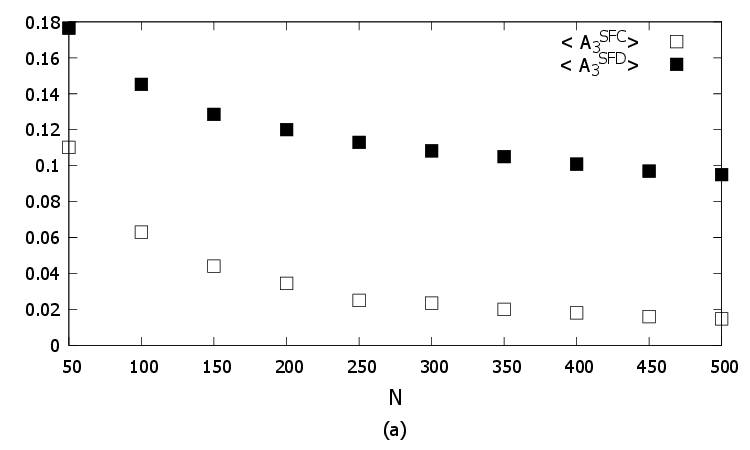}
		\includegraphics[width=100mm]{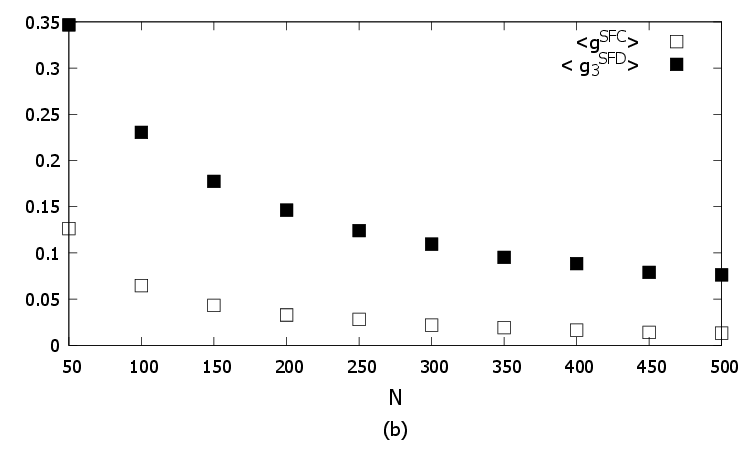}
	\end{center}
	\caption{Mean asphericities $\langle A_3\rangle$ (a) and size ratios $\langle g \rangle$ (b) of polymer scale-free networks with fixed $\alpha=3.5$ and various $N$ obtained in numerical simulations with application of Wei formula (\ref{gwei}). Open symbols: SFC algorithm; filled symbols: SFD algorithm.  }
	\label{AN}
\end{figure}

\subsection{Results for SFC and SFD networks}\label{IIId}
We have constructed the ensembles of $M=1000$ 
networks applying both SFC and SFD schemes, and considered networks of the size $N$ in the range up to 500. Let us remind that in each case the averaging of all the observables over an ensemble of constructed networks  is performed according to (\ref{aver}).

The node-degree distributions $P(k)$ in SFC and SFD networks at fixed $N=100$ and $\alpha=3.5$ are presented in Fig.  \ref{kser}a. {    Let us remind, that the probability
to create a node with $k=1$ is zero during the process of SFD network construction, but when the construction stops the
networks will contain  the
peripheral nodes with degree $1$. Due to the presence of these open nodes, the probabilities of obtaining nodes with degrees $k>1$ are slightly shifted for the case of SFD model as comparing with SFC, due to the highest probability of $P(k=1)$ in this case. 
The slope of both curves in a double logarithmic scale is the same according to the power law (\ref{dis}).
The intrinsic difference between networks constructed according to  SFC and SFD  algorithms affects also the averaged degree  value $k_{av}$  calculated on the basis of node-degree distribution $P(k)$ in both cases, as presented in Fig. \ref{kser}b for different $\alpha$. For the SFC network, $k_{av}$ decreases rapidly with increasing $\alpha$ due to lower probabilities of obtaining the high degree nodes. For the SFD network, the dependence of $k_{av}$ on $\alpha$ is more smooth. It approaches the value of two even for networks with very small $\alpha$ due to the major role played by nodes with degree one,   similarly as it takes place for regular dendrimer structures considered above (cf. Eq. \ref{kavdend}). }

Again, let us start with considering  the peculiarities of eigenvalue spectra $\sigma_i$ of Kirchhoff matrix of such complex structures. In Fig. \ref{EigenNet}a, we present our data  for $\sigma_i$ of the networks of $N=100$ nodes with several values of $\alpha$, constructed on the basis of SFD algorithm, in comparison with those of simple linear chain. We observe, that with increasing $\alpha$ the spectrum of scale-free dendrimer structures tends to coincide with those of the linear chain, supporting the intuitive expectation that such structures tend to degenerate into a regular chain in the limit of large $\alpha$. On the other hand, the smaller is $\alpha$, the larger values of $\sigma_i$ are observed, signalizing that such networks are more compact and symmetric as compared to the linear structure.

  An estimate for the upper bound for the eigenvalue spectrum of Kirchhoff matrix according to (\ref{maxeig}) can be obtained, taking into account the following consideration. Let us recall, that the  nodes with degree $k$ are observed in a network with probability $P(k)$, depending also on the size $N$ of a network. At small $N$, probability of observing the node with large values of $k$ is in fact negligibly small. We can introduce the value $n(k,N)=p(k,N)\times N$, as a number of nodes with degree $k$ actually present in a structure with total number of nodes $N$ (see Fig. \ref{nnet}a). The node with a larger possible degree, satisfying (\ref{maxeig}), is the largest possible, satisfying condition $\langle n(k,N)\rangle\geq 1$. E.g. for the case $\alpha=5.5$, that condition is satisfied at $k=4$, giving $\sigma_{{\rm sup}}=8$, whereas at $\alpha=10.5$ -- at $k=3$, thus $\sigma_{{\rm sup}}=6$. Note that at fixed value of $\alpha$, the values of $n(k,N)$ increase with increasing the size of network (Fig. \ref{nnet}b), signalizing, correspondingly, a gradual increase of maximal possible eigenvalues with $N$.  

The size and shape characteristics of both types of structures are addressed next. 
Results for $\langle A_3\rangle$ and $\langle g \rangle$, obtained by applying the Wei's formulas (\ref{awei}), (\ref{gwei}) and performing averaging over an ensemble of constructed networks at fixed $N=100$ and variable $\alpha$ are presented in Fig. \ref{A100}. 
Both values are increasing with $\alpha$, indicating the decreasing role of nodes with high degree, while at any $\alpha$ the SFD structures are more anisotropic and extended as compared to the  SFC constructed networks. Note that for sufficiently high $\alpha$ the SFC networks tend to the structure of a circular polymer ring, while the SFD network would reach the limit of linear polymer chain. For these cases, the known values $A_{3 {\rm chain}}=0.394274$ \cite{Wei1997a}, $A_{3 {\rm ring}}=0.246368$ \cite{Wei1997a}, $g_{ {\rm ring}}=1/2$ \cite{Zimm1949}
are approached.

The dependence of size and shape parameters on the network size $N$ is presented in Fig. \ref{AN}. 
For finite network size $N$, the values of shape parameters differ from those for infinite systems. This finite-size deviation obeys the scaling behavior with $N$:
\begin{eqnarray}
\langle A_3(N) \rangle = \langle A_3 \rangle+B/N^{\delta},\\
\langle g(N) \rangle = \langle g \rangle+C/N^{\delta}, \label{fit}
\end{eqnarray}
with $B$, $C$ being some constants and $\delta$ is the correction-to-scaling exponent $\delta=3/2$ \cite{Caracciolo2005}. The results obtained by applying the least-square fitting of the data to the form (\ref{fit}) are given in Table 1. 
Note that these limiting values do not approach the expected values of chain and ring correspondingly even at considerably high values of $\alpha$.

\label{tab}
\begin{table}[b!]
\begin{center} {\small{
    \begin{tabular}{|c|c|c|c|c|c|c|}
    \hline 
     $\alpha$ &   $k_{av}^{SFC}$ & $\langle A_3 ^{SFC} \rangle$  & $\langle g^{SFC} \rangle$  & $k_{av}^{SFD}$ &$\langle A_3^{SFD} \rangle$  &$\langle g^{SFD} \rangle$ \\ \hline
2.5 & 4.721 & 0.004(4) & 0.006(5)  &2.029 & 0.068(2)  &0.037(4) \\
3.5 &  2.694   &0.012(4) &0.016(5) & 1.999 & 0.094(3) &0.069(5) \\
4.5  & 2.316  &0.019(4) &0.026(5) &1.998  & 0.114(4) &0.114(8) \\
5.5 &  2.170  &0.041(5)  &0.051(6)  & 1.997  &0.137(5) &0.169(8) \\
6.5   & 2.099  &0.067(6) &0.095(6) &1.997  & 0.159(7) &0.24(1)  \\
7.5 &  2.060  &0.103(6) &0.134(8)  &1.997 & 0.187(8) &0.32(1)  \\
8.5 &  2.037 &0.149(9) & 0.235(8) & 1.997   & 0.216(9) & 0.41(1)  \\
9.5  & 2.024 &0.190(9)&0.305(9) &1.996 & 0.25(1) &0.52(1)  \\
10.5 & 2.015 & 0.220(9) & 0.352(9) &1.996 &0.27(1) &0.63(1)\\ \hline       
          \end{tabular}
          }}
    \caption{Results of the least-square fits for evaluation of the asymptotic values for $k_{av}$, $\langle A_3\rangle$, $\langle g \rangle$ in the limit $N\to \infty$ of the networks constructed on the basis of SFC and SFD algorithms.}
    \end{center}
        \end{table}


To shed more light on the observed behaviour, let us recall that although the probabilities to obtain the branching points (nodes with degrees higher than two) are very small at large $\alpha$ according to (\ref{dis}), they increase with growing the network size $N$ (cf. Fig. \ref{nnet}). Thus, in the limit  $N\to\infty$  there is still significant (yet very small) number of such nodes appearing in a network even at very large $\alpha$.  Thus, though in this limit the resulting networks tend to form regular linear (at $k_{{\rm min}}=1$) or circular (at $k_{{\rm min}}=2$) structures, correspondingly, the not-negligible probability of occurring the branching points with $k>2$ holds even at large $\alpha$. This explains the deviation of our results for size and shape characteristics, presented in Table  1  from the expected values for linear and circular polymer structures, as it is observed for smaller network size (cf. Fig. \ref{A100}).

\section{Conclusions}\label{IV}

In the present study, we analyzed an impact of topology of the complex hyperbranched polymer structures on their conformational properties in Gaussian regime. 
Idealized Gaussian chains considered in our analysis serve as a standard approximation in polymer theory. Being simple enough to allow for an analytical
treatment, in many cases Gaussian chain approximation allows to shed light on a non-trivial behaviour of polymers and their agglomerations. A classical example is given by Werner Kuhn's treatment of a random walk shape \cite{Kuhn1934},
showing that a non-spherical shape is inherent even to the polymer coil formed by a self-intersecting Gaussian chain and explaining experimentally observed
rheological properties of polymer macromolecules in a solvent. Despite being considered as a formal step towards analysis effects of a self-avoidance, Gaussian chains are also experimentally realizable in the so-called theta-solvents, when the
self-avoidance effectively vanishes due to compensation of attraction and repulsion 
terms in the second virial coefficient \cite{desCloiseaux}.

 In the mathematical language of  graphs, the branching points in hyperbranched networks can be treated as
vertices (nodes), and their functionalities as degrees of these nodes. Following
  Refs. \cite{Galiceanu2014,Galiceanu2012}, we turned our attention to the scale-free polymer networks with functionalities $k$ (the degrees) of nodes obeying a power-law  distribution (\ref{dis}). We used two types of approaches, allowing to construct such types of networks, namely the one based on configuration model algorithm \cite{Bender1978,Molloy1995}  (assuming $k_{{\rm min}}=2$), and the scale-free dendrimer  algorithm \cite{Galiceanu2014,Galiceanu2012} (with $k_{{\rm min}}=1$). Note that the latter one can be treated as generalization of regular hierarchical denrimer structure, where the node degrees are fixed (cf. Figs. \ref{snapdend} and \ref{snapnet}).

It is interesting to note that when considering the processes of ordering on scale-free networks, it is the decay exponent $\alpha$ of the node-degree distribution
$P(k)$, Eq. (1), that defines the universality class and even the possibility of a
phase transition \cite{Dorogovtsev2008}.
Similarly, when considering the stability of networks, the Molloy-Reed criterion  \cite{Molloy1995,Molloy1998} predicts different stability of scale-free networks depending on the decay exponent. For example,
scale-free networks with $\alpha<3$ always contain giant connected component \cite{Cohen2002}.
In this sense, the problem considered by us in this paper is an interesting example of
how different scale-free networks with the same value of $\alpha$ nevertheless have
different universal shape characteristics.
  
To evaluate the numerical values for the observables of interest, such as the size ratio  (as defined by Eq. (\ref{gratio}) and asphericity (\ref{asfer}) of complex scale-free polymer structures, we made use of the  Wei's method \cite{Wei,Ferber2015}, which is based on evaluation of the eigenvalue spectra of Kirchhoff matrices of corresponding graphs. Already on the level of analysis of eigenvalues spectra, we qualitatively conclude an increase of compactness and sphericity of shape of such polymer structures  for small $\alpha$, whereas at large enough  $\alpha$
the structures have the tendency to convert into the linear chain (for the case $k_{{\rm min}}=1$) or circular ring (when $k_{{\rm min}}=2$), respectively.
We constructed the ensembles of $1000$ 
networks applying the both algorithms, and considered networks of the size $N$ in the range up to 500 in a wide range of parameter $\alpha$.
Separately we considered the case of small network size $N=100$.  Both $\langle A_3\rangle$ and $\langle g \rangle$ values in this case are increasing with $\alpha$, indicating the decreasing role of branching points  with high degree, while at any $\alpha$ the SFD structures are more anisotropic and extended as comparing with SFC constructed networks.  The results obtained by us for infinite networks by applying the least-square fitting of our data  are summarized in Table 1. The  significant  number of branching points (nodes with $k>2$)  appearing in  infinite networks   explains the deviation of resulting values even at very large $\alpha$ from the expected values for linear and circular polymer structures.

\section*{Acknowledgement} We are indebted to Dr. Christian von Ferber for  inspiring us to investigate the present topic. 

\section*{References}
\bibliographystyle{iopart-num} 
\bibliography{shape}

\end{document}